\documentclass[oupdraft]{bio}
\usepackage{url, bm, bbm, changepage}

\newcommand{\expect}{\mathbbm{E}}
\newcommand{\indic}[1]{\mathbbm{1}\left[{#1}\right]}
\newcommand{\expit}{\mbox{expit}}

\history{Received XXX; revised XXX; accepted for publication XXX}

\begin{document}

\title{A joint Bayesian hierarchical model for estimating SARS-CoV-2 diagnostic and subgenomic RNA viral dynamics and seroconversion}

\author{Tracy Q. Dong$^{1\ast}$ and Elizabeth R. Brown$^{1,2}$ \\[4pt]
$^{1}$\textit{Vaccine and Infectious Disease Division, Fred Hutchinson Cancer Center} \\
$^{2}$\textit{Department of Biostatistics, University of Washington}
\\[2pt]
{qdong@fredhutch.org}}

\markboth%
{T.Q. Dong and E.R. Brown}
{Joint Bayesian model for SARS-CoV-2 viral load and seroconversion}

\maketitle

\footnotetext{To whom correspondence should be addressed.}

\begin{abstract}
{Understanding the viral dynamics and immunizing antibodies of the severe acute respiratory syndrome coronavirus 2 (SARS-CoV-2) is crucial for devising better therapeutic and prevention strategies for COVID-19. Here, we present a Bayesian hierarchical model that jointly estimates the diagnostic RNA viral load reflecting genomic materials of SARS-CoV-2, the subgenomic RNAs (sgRNA) viral load reflecting active viral replication, and the rate and timing of seroconversion reflecting presence of antibodies. Our proposed method accounts for the dynamical relationship and correlation structure between the two types of viral load, allows for borrowing of information between viral load and antibody data, and identifies potential correlates of viral load characteristics and propensity for seroconversion. We demonstrate the features of the joint model through application to the COVID-19 PEP study and conduct a cross-validation exercise to illustrate the model's ability to impute the sgRNA viral trajectories for people who only had diagnostic viral load data. }
{Joint models; Bayesian hierarchical models; SARS-CoV-2; Correlates; Viral load; Seroconversion.}
\end{abstract}

\section{Introduction} \label{sec_intro}

The severe acute respiratory syndrome coronavirus 2 (SARS-CoV-2) has spread rapidly around the world, resulting in more than 6 million deaths worldwide to date (\citealp{who2022covid}). Current knowledge regarding SARS-CoV-2 viral dynamics is built on diagnostic viral load detected via real-time reverse transcription polymerase chain reaction (RT-PCR)(\citealp{cevik2021sars, gastine2021systematic, mellis2022sars, pajon2022initial}), the most sensitive and widely used technique for the diagnosis of coronavirus disease 2019 (COVID-19). However, the diagnostic RT-PCR tests give limited insight into infectivity because the results may reflect residual genomic material rather than replicating virus (\citealp{singanayagam2020duration}). Viral culture and quantification by plaque assays provide a gold standard for assessing infectivity but are not readily available at large scale for a diagnostic assay. Therefore, detection of subgenomic RNAs (sgRNA) in clinical samples has been suggested as an additional diagnostic tool to track infectious virus because sgRNA transcripts indicate active viral replication (\citealp{dagotto2021comparison, santos2022viral}). Characterization of both SARS-CoV-2 diagnostic and sgRNA viral dynamics is crucial for understanding the pathogenesis of the virus, identifying patients most at risk , and optimizing the period of self-isolation to reduce onward transmission while reducing unnecessary loss of productivity. 

Developing a model for SARS-CoV-2 viral load is of particular interest because viral load is an important outcome of COVID-19 treatment, vaccine, and non-vaccine prevention studies but viral load trajectories are rarely fully observed (\citealp{levine202a1initial, gottlieb2021effect, goldberg2021real, chaccour2021effect, portal2022virologic, levine2022waning, eyre2022effect}). One commonly used approach is compartmental models based on ordinary differential equations (ODE) (\citealp{goyal2020potency, neante2020modeling, ke2021vivo, challenger2022modelling}). While the ODE-based models are useful for capturing cell-level within-host viral kinetics and simulating potential effects of antiviral therapies under various assumptions, the model parameters cannot be easily translated into key viral trajectory characteristics such as peak viral load and shedding duration. Therefore, Bayesian mixed-effects models are a popular alternative for estimating population- and individual-level viral load trajectories (\citealp{kissler2021viral, kissler2021viral1, SK2021trajectory, jones2021estimating, singanayagam2021community, deming2022Detection}). This approach has the advantage of being able to directly model viral load features and incorporate covariates and correlation structures to explore potential associations between viral dynamics and patient characteristics. So far, all existing models estimate the diagnostic and sgRNA viral load separately without considering any dependency between the two. 

In addition to viral dynamics, serology characteristics of SARS-CoV-2 infection, especially the rate and timing of seroconversion defined by the development of neutralizing antibodies, are also of interest. Not all persons recovering from SARS-CoV-2 infection develop antibodies (\citealp{oved2020multi, pathela2021seroprevalence, wellinghausen2020sars}). Understanding the natural immunity acquired after SARS-CoV-2 infection and the correlates of seroconversion is crucial for studying disease transmission, vaccine effectiveness, and population-wide seroprevalence and immunity. Since most serological studies were conducted among hospitalized people who exhibited moderate to severe COVID-19 symptoms and were tracked in the later course of the infection (\citealp{lou2020serology, yadav2021seroconversion, masia2021sars}), little is known regarding seroconversion rate and time in outpatients settings and among persons with very mild or asymptomatic infection (\citealp{cdc2021science}). 

In this paper, we develop a Bayesian hierarchical model that jointly analyzes SARS-CoV-2 diagnostic and sgRNA viral load data and interval-censored seroconversion data to estimate viral trajectories, their association with patient characteristics, and the correlates, rate, and time of seroconversion. Our proposed model has several advantages compared to the existing models: first, it takes into account the dynamical relationship between the diagnostic PCR and sgRNA viral trajectories and can be used to impute sgRNA viral trajectories for patients who only had diagnostic viral load data. In addition, by jointly modeling viral load and antibody measurements, our model allows the borrowing of information across different data types to better estimate key virological and serological parameters. Our model also uses a pair of indicator variables for the observed and true viral shedding status to allow direct modeling of false positive and false negative cases. Finally, our model explicitly accounts for the increased uncertainty in RT-PCR viral load measurements below the test's limit of quantification (LoQ). 

In Section \ref{sec_example}, we introduce a motivating example of the COVID-19 post-exposure prophylaxis (PEP) study (\citealp{barnabas2021hydroxychloroquine}). Section \ref{sec_methods} describes the model formulation and inference procedure. In Section \ref{sec_application}, we apply our methods to analyze data collected in the COVID-19 PEP study and summarize the results. We end the paper with a discussion in Section \ref{sec_discuss}.

\section{A motivating example: the COVID-19 PEP Study} \label{sec_example}

The COVID-19 PEP Study was a double-blinded, household-randomized controlled trial comparing hydroxychloroquine to placebo-like control for prevention of SARS-CoV-2 infection (\citealp{barnabas2021hydroxychloroquine}). The study was remotely conducted between March and August 2020 before COVID-19 vaccines were available and enrolled 829 initially asymptomatic household contacts and healthcare workers recently exposed ($<$96 hours) to persons with laboratory-confirmed SARS-CoV-2 infection from 41 U.S. states. The study found no differences between hydroxychloroquine and placebo-like control arms in SARS-CoV-2 acquisition. 

Participants provided demographic information via enrollment surveys. They also completed daily symptom questionnaires and self-collected daily mid-turbinate swabs for 14 consecutive days. Swabs were tested for SARS-CoV-2 RNA at the University of Washington (UW) Virology Laboratory using RT-PCR targeting nucleocapsid genes N1 and N2. Samples with detectable SARS-CoV-2 RNA via RT-PCR were further tested for sgRNA targeting sgE (\citealp{bruce2021predicting}). 

RT-PCR for both N and sgE were performed to quantify the viral titer via the cycle threshold (Ct), defined as the number of thermal cycles needed to amplify sampled viral RNA to a detectable level. Quantification standards were run with both assays to convert Ct value to viral load as copies/ml during assay validation. Both tests used a Ct cutoff of 40, corresponding to a lower limit of detection (LoD) of approximately $10^{2.4}$ copies/ml. In addition, the diagnostic RT-PCR had a limit of quantification (LoQ), defined as the lowest amount of viral RNA in a sample that can be quantitatively determined with acceptable precision accuracy (\citealp{forootan2017limit}), of approximately $10^{4.9}$ copies/ml, corresponding to a Ct value of 31. Therefore, the conversion from the diagnostic Ct values to viral concentrations suffers higher uncertainty for nasal swabs with observed Ct values between the LoD and LoQ. 

Dried blood spot (DBS) samples were voluntarily provided by a subset of participants on study day 1, 14 and 28. The DBS samples were assayed to detect Immunoglobulin G (IgG) antibodies to SARS-CoV-2 spike protein at the UW Department of Laboratory Medicine \& Pathology. For brevity, we will use ``seroconversion" to refer to anti-spike IgG seroconversion for the rest of the paper. However, the model can be generalized to accommodate other types of seroonversion as well. More information can be found in the Discussion section. 

\section{Methods}\label{sec_methods}

\begin{figure}[!p]
	\centering
	\includegraphics[width=1\textwidth]{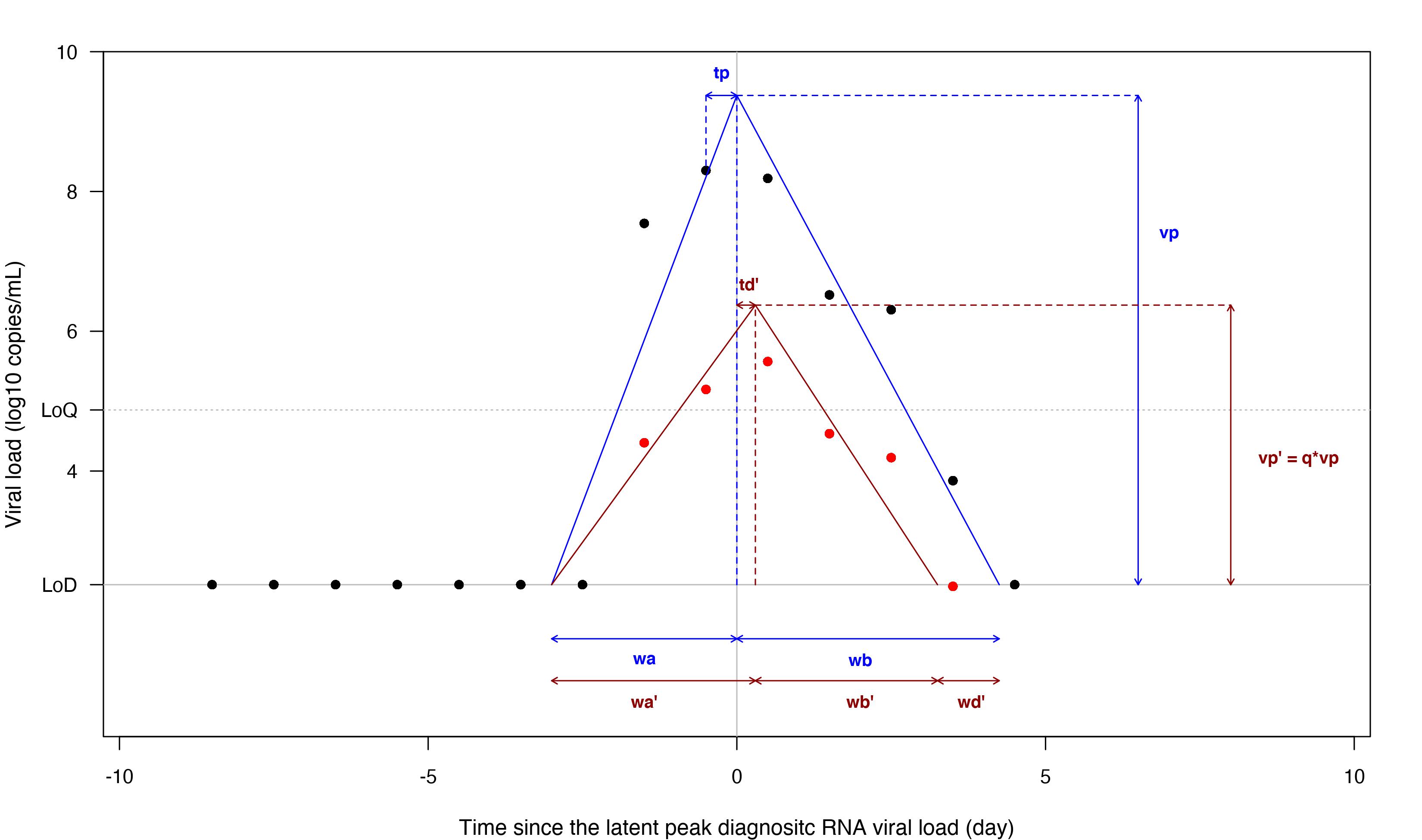}
	\caption{Graphical illustration of the joint model. Along the x-axis is time since the latent peak diagnostic RNA viral load in day. Along the y-axis is the magnitude of viral load in log10 copies/ml. The limit of detection (LoD) for both diagnostic and sgRNA RT-PCR tests were approximately $10^{2.4}$ copies/ml. In addition, the diagnostic RT-PCR had a limit of quantification (LoQ) of approximately $10^{4.9}$ copies/ml. Black dots represent the observed diagnostic RNA viral load from RT-PCR tests, and blue segments represent the latent diagnostic RNA viral load trajectory. Red dots represent the observed sgRNA viral load from RT-PCR tests, and maroon segments represent the latent sgRNA viral load trajectory. The black and red dots were actual data from one participant from the COVID-19 PEP study. $w_a$ and $w_a'$ represent time from shedding onset to peak for diagnostic and sgRNA viral load respectively. $w_b$ and $w_b'$ represent time from peak to viral clearance for diagnostic and sgRNA viral load respectively. $w_d'$ represents the time from sgRNA viral clearance to diagnostic RNA viral clearance. $t_p$ represents time from the observed peak diagnostic RNA viral load to the latent peak diagnostic RNA viral load, and $t_d'$ represents time from the latent peak diagnostic RNA viral load to the latent peak sgRNA viral load. Finally, $v_p$ represents the magnitude of peak diagnostic RNA viral load with respect to LoD, and $q$ is the multiplicative factor of peak sgRNA viral load relative to peak diagnostic RNA viral load. }
	\label{fig_diagram}
\end{figure}

We assume that the diagnostic and sgRNA viral load trajectories consist of a proliferation phase with exponential growth in viral RNA concentration, followed by a clearance phase characterized by exponential decay in viral RNA concentration. On the scale of log10 copies/ml, this roughly corresponds to a linear increase to a peak followed by a linear decrease. Figure \ref{fig_diagram} shows a graphical illustration of the latent viral RNA concentration trends overlay with the observed diagnostic and sgRNA viral load measurements over time for an individual participant. We describe the details of the model in the following sections.

\subsection{Modeling the diagnostic RNA viral load} \label{sec_diagRNA}

Let $i$ be index for participants and $j$ be index for nasal swabs. For person $i$, let $w_{a,i}$ represent time from shedding onset to peak viral load, $w_{b,i}$ represent time from peak to viral clearance, and $v_{p,i}$ represent the magnitude of peak viral load with respect to LoD (Figure \ref{fig_diagram}). We use the latent peak as a reference point and let $s$ represent time since the latent peak viral load. As such, we have $s = 0$ at the latent peak viral load, $s = -w_{a,i}$ at shedding onset, and $s = w_{b,i}$ at viral clearance. For the $j$th swab collected from person $i$, $s_{ij}$ is the time from peak viral load to swab collection, and the latent diagnostic viral load at $s_{ij}$, denoted by $\mu_{y} (s_{ij})$, is \begin{align*}
    \mu_{y} (s_{ij}) =  \mbox{LoD} + v_{p,i} + \left( \frac{v_{p,i}}{w_{a,i}}  \right) \times s_{ij} \times \indic{ s_{ij} \le 0 } + \left( -\frac{v_{p,i}}{w_{b,i}}  \right) \times s_{ij} \times \indic{ s_{ij} > 0 }, 
\end{align*}
where $\indic{\cdot}$ is an indicator function. In particular, $\left( \frac{v_{p,i}}{w_{a,i}} \right)$ and $\left( -\frac{v_{p,i}}{w_{b,i}} \right)$ represent the upward and downward slopes respectively for the linear change in diagnostic viral trajectory over time. To model the person-level variability in $w_{a,i}$, $w_{b,i}$ and $v_{p,i}$ while accounting for potential correlation among these parameters, we use a multivariate normal distribution for the random effects on the log scale: 
\begin{align*}
\log\left( [v_{p, i}, w_{a, i},  w_{b, i} ]^{\intercal}  \right) \sim &~ \mbox{MVN}\left( \bm{\mu}_{log,i}, \bm{\Sigma}_{log} \right) \\
\bm{\mu}_{log, i}[1] = &~ \mu_{lvp} + \bm{\beta}_{vp}^{\intercal} \bm{X}_{vp, i} \\
\bm{\mu}_{log, i}[2] = &~ \mu_{lwa} + \bm{\beta}_{wa}^{\intercal} \bm{X}_{wa, i} \\
\bm{\mu}_{log, i}[3] = &~ \mu_{lwb} + \bm{\beta}_{wb}^{\intercal} \bm{X}_{wb, i} 
\end{align*}
where $\bm{X}_{vp, i}$, $\bm{X}_{wa, i}$ and $\bm{X}_{wb, i}$ are vectors of person-level covariates for peak viral load, time from shedding onset to peak, and time from peak to viral clearance, respectively; and $\bm{\beta}_{vp}$, $\bm{\beta}_{wa}$ and $\bm{\beta}_{wb}$ are the corresponding vectors of coefficients for these covariates. 

We use a pair of indicator variables to model the true positive rate (TPR) and true negative rate (TNR) for the observed diagnostic viral detection. Let $y_{ij}$ denote the observed viral RNA concentration in log10 copies/ml for swab $j$ of person $i$, $B_{ij}$ be the fully observed indicator variable of whether the swab had detectable diagnostic RNA via RT-PCR (i.e., $B_{ij} = \indic{y_{ij} > \mbox{LoD}}$), and let $S_{ij}$ be the latent indicator variable of whether person $i$ was truly experiencing diagnostic viral shedding when swab $j$ was collected (i.e., $S_{ij} = \indic{ -w_{a,i} \le s_{ij} \le w_{b,i}}$). We specify the following model:
\begin{align*}
	B_{ij} | S_{ij} \sim &~\mbox{Bernoulli} \left( \expit ( \alpha_0 + \alpha_1 S_{ij} ) \right).
\end{align*}
In particular, the true positive rate (TPR), defined as the probability of detecting diagnostic RNA from a nasal swab when a person was experiencing viral shedding, is 
\begin{align*}
\mbox{TPR} = \expect [B_{ij} = 1 | S_{ij} = 1] = \expit(\alpha_0 + \alpha_1);
\end{align*}
and the true negative rate (TNR), defined as the probability of not detecting viral RNA from a nasal swab when a person was clear of viral RNA, is 
\begin{align*}
\mbox{TNR} = \expect [B_{ij} = 0 | S_{ij} = 0] = 1 - \expit(\alpha_0).
\end{align*}

The distribution of a swab's observed diagnostic viral load depends on whether the test result is a true/false positive/negative. When the $j$th swab collected from person $i$ has no detectable diagnostic RNA, i.e., $B_{ij} = 0$, the observed viral load equals to the LoD. When a swab does have detectable viral RNA, i.e., $B_{ij} = 1$, we further consider two scenarios: if the result is false positive, i.e., $S_{ij} = 0$, we use a normal distribution centered around some viral load level above the LOD with a moderate standard deviation:
\begin{align*}
y_{ij} | B_{ij} = 1, S_{ij} = 0 \sim \mbox{Normal} \left( \mbox{LoD}^{+}, 0.5 \right).
\end{align*}
If the result is true positive, i.e., $S_{ij} = 1$, we use a normal distribution centered around the latent viral trajectory and a standard deviation that depends on LoQ:
\begin{align*}
y_{ij} | B_{ij} = 1, S_{ij} = 1 \sim \mbox{Normal} \left( \mu_{y} (s_{ij}), \sigma_{y} \left( Q_{ij} \right) \right),
\end{align*}
where 
\begin{align*}
    \sigma_{y}^2 \left( Q_{ij} \right) = \sigma^2_{yy} \times \left(1 + Q_{ij} \times \delta_{Q} \right).
\end{align*}
Our specification the variance $\sigma_{y}^2$ aims to account for the increased uncertainty in viral load measurements for swabs with diagnostic RNA concentration below the LoQ. Specifically, we let $\sigma^2_{yy}$ be the base variance term denoting the unmeasured variability in the observations, and $Q_{ij}$ be an indicator variable of whether the observed viral concentration from swab $j$ lies between LoD and LoQ (i.e., $Q_{ij} = \indic{ \mbox{LoD} < y_{ij} < \mbox{LoQ}}$). When a swab has an observed diagnostic RNA viral load between LoD and LoQ, i.e., $Q_{ij} = 1$, the variance term will be higher than the base level by a factor of $\delta_Q$ to account for the extra uncertainty in these observations. 

Finally, since the latent peak viral load is not observable, we estimate the time from the observed peak to the latent peak for person $i$, denoted $t_{p,i}$. If person $i$'s observed peak viral load was recorded within the first two days of follow-up, we will assume the latent peak happens at the latest one day after the observed peak:
\begin{align*}
	t_{p, i} \sim \mbox{Normal} \left(\mu_{tpL}, \sigma_{tpL} \right) \mbox{truncated }[-\infty, 1],
\end{align*}
where $\mu_{tpL}$ is negative. If person $i$'s observed peak viral load was recorded during the last two days of follow-up, we will assume the latent peak happens at the earliest one day before the observed peak:
\begin{align*}
	t_{p, i} \sim \mbox{Normal} \left(\mu_{tpR}, \sigma_{tpR} \right) \mbox{truncated }[-1, \infty],
\end{align*}
where $\mu_{tpR}$ is positive. Otherwise, we assume the latent peak happens around the time of the observed peak:
\begin{align*}
	t_{p, i} \sim \mbox{Normal} \left( 0, \sigma_{tp} \right).
\end{align*}
We can specify $\sigma_{tpL}$, $\sigma_{tpR}$, and $\sigma_{tp}$ to reflect our prior belief in how variable the time differences $t_{p, i}$ are among the participants. 

\subsection{Modeling the sgRNA viral load} \label{sec_sgRNA}

We model the sgRNA viral load in relation to the diagnostic viral trajectory. For clarity, we define a parallel set of notations for the sgRNA viral load by adding $'$ after the previously defined notations. We assume that the diagnostic and sgRNA viral shedding starts at the same time. In addition, we know that sgRNA viral clearance occurs no later than the diagnostic RNA viral clearance (Figure \ref{fig_diagram}). 

Let $t_{d,i}$ denote time from the latent peak diagnostic viral load to the latent peak sgRNA viral load for person $i$. We use the following truncated normal distribution for $t'_{d,i}$:
\begin{align*}
	t'_{d,i} \sim &~ \mbox{Normal} \left( \mu_{td}, 2  \right) \mbox{ truncated }  [-w_{a,i}, w_{b,i}] 
\end{align*}
The time from sgRNA shedding onset to peak viral load for person $i$, denoted $w'_{a,i}$, can be calculated as
\begin{align*}
	w'_{a,i} = w_{a,i} + t'_{d,i}.
\end{align*}
We use a truncated normal distribution to model the log-transformed time from sgRNA viral clearance to diagnostic RNA rival clearance, denoted $w'_{d, i}$:
\begin{align*}
   \log \left( w'_{d, i} \right) \sim  \mbox{Normal} \left(\mu_{lwd},  \sigma_{lwd} \right) \mbox{ truncated }  \left[0, \log \left( w_{b,i} - t'_{d,i} \right) \right]. 
\end{align*} 
The time from peak sgRNA viral load to viral clearance for person $i$, denoted $w'_{b,i}$, can be calculated as 
\begin{align*}
	w'_{b,i} = w_{a,i} + w_{b,i} - w'_{a,i} - w'_{d,i}.
\end{align*} 

As for the magnitude of the peak sgRNA viral load $v'_{p,i}$, we parameterize it as the product of the diagnostic peak viral load and a multiplier $q_i$, i.e., $ v'_{p,i} = q_i \times v_{p,i} $. Since the peak sgRNA viral load is always lower than the peak diagnostic RNA viral load, we model the multiplier $q_i$ using a beta distribution:
\begin{align*}
	q_i \sim \mbox{Beta}\left(\gamma_1, \gamma_2 \right)
\end{align*} 

In the PEP study, only the swabs with detectable diagnostic RNA were tested for sgRNA. Therefore, we model the TPR and TNR of sgRNA detection conditional on the observed diagnostic viral load measurement: 
\begin{align*}
	B'_{ij} | S'_{ij}, B_{ij} \sim &~\mbox{Bernoulli} \left( B_{ij} \times \left[  \expit (\alpha'_0 + \alpha'_1 S'_{ij} ) \right]  \ \right).
\end{align*}
Specifically, when a swab was negative for diagnostic RNA, i.e., $B_{ij} = 0$, it would also be negative for sgRNA, i.e., $B'_{ij} = 0$. Otherwise, the probability of detecting sgRNA from a nasal swab when a person was experiencing sgRNA viral shedding is 
\begin{align*}
\mbox{TPR}' = \expect [B'_{ij} = 1 | S'_{ij} = 1, B_{ij} = 1] = \expit(\alpha_0' + \alpha_1');
\end{align*}
and the probability of not detecting sgRNA from a nasal swab when a person was clear of sgRNA is 
\begin{align*}
\mbox{TNR}' = \expect [B'_{ij} = 0 | S'_{ij} = 0, B_{ij} = 1] = 1 - \expit(\alpha_0').
\end{align*}

The distribution of the observed sgRNA viral load follows a similar structure as the diagnostic viral load and depends on whether the sgRNA test result is a true/false positive/negative. When the $j$th swab collected from person $i$ has no detectable sgRNA, i.e., $B'_{ij} = 0$, the observed viral load equals to the LoD. When a swab does have detectable sgRNA and the result is false positive, we use a normal distribution centered around some viral load level above the LOD with a moderate standard deviation:
\begin{align*}
y'_{ij} | B'_{ij} = 1, S'_{ij} = 0 \sim \mbox{Normal} \left( \mbox{LoD}^{+}, 0.5 \right).
\end{align*}
Finally, if the sgRNA viral load detected is true positive, we use a normal distribution:
\begin{align*}
y'_{ij} | B'_{ij} = 1, S'_{ij} = 1 \sim \mbox{Normal} \left( \mu_{y}' (s'_{ij}), \sigma'_{y} \right),
\end{align*}
where 
\begin{align*}
    \mu'_{y} (s'_{ij}) =  \mbox{LoD}' + v'_{p,i} + \left( \frac{v'_{p,i}}{w'_{a,i}}  \right) \times s'_{ij} \times \indic{ s'_{ij} \le 0 } + \left( -\frac{v'_{p,i}}{w'_{b,i}}  \right) \times s'_{ij} \times \indic{ s'_{ij} > 0 }.
\end{align*}

\subsection{Modeling seroconversion rate and time} \label{sec_sero}

Since not everyone infected with SARS-CoV-2 would eventually develop neutralizing antibodies, we use an indicator variable $C_i$ to model whether person $i$ would ever seroconvert: 
\begin{align*}
	C_i \sim \mbox{Bernoulli} \left( \expit \left( \beta_{C0} + \bm{\beta}_{C}^{\intercal} \bm{X}_{C, i} \right)  \right),
\end{align*}
where $\bm{X}_{C, i}$ are vectors of potential correlates of seroconversion. If a person developed antibodies after infection, then the time from viral shedding onset to seroconversion, denoted $w_{s, i}$, is modeled as 
\begin{align*}
	w_{s, i} | C_i = 1 \sim &~ \mbox{Gamma}\left( \kappa_1, \kappa_2  \right) 
\end{align*}
If a person never seroconverts, $w_{s, i} = \infty$. 

In our motivating example of the COVID-19 PEP study, DBS samples were only collected on day 1, 14 and/or 28 of the study. Therefore, the servoconversion time is censored. Specifically, if a participant only had negative IgG samples, his/her seroconversion time is right-censored at the last negative antibody test; if a participant only had positive IgG samples, his/her seroconversion time is left-censored at the first positive antibody test; if a participant had both negative and positive IgG samples, his/her seroconversion time is interval-censored between the last negative test and the first positive test. Finally, since not all participants had provided DBS samples with confirmed results, only those who had antibody data available will contribute to the likelihood of the model. 

\subsection{Prior elucidation} \label{sec_prior}

Inference for all model parameters can be conducted under a Bayesian framework. The prior distributions for virological parameters $\left\{ \mu_{lvp}, \mu_{lwa}, \mu_{lwb}, \mu_{lwd} \right \}$ can be specified using normal distributions with means equal to the estimated diagnostic and sgRNA peak viral load, time from shedding onset to peak and time from peak to shedding cessation from previous literature. For variance and covariance parameters $\left\{ \sigma_{yy}^2, \sigma_{y}^{'2}, \bm{\Sigma}_{log} \right \}$, we can use weakly informative priors such as the Cauchy distribution and inverse wishart distribution. Multivariate normal distributions with zero means can be used as priors for the coefficients of potential correlates of viral dynamics and seroconversion $\left\{ \bm{\beta}_{vp}, \bm{\beta}_{wa}, \bm{\beta}_{wb}, \bm{\beta}_{C} \right \}$. For the model parameters related to TPR and TNR of diagnostic and sgRNA detection $\left\{ \alpha_0, \alpha_1, \alpha_0', \alpha_1' \right \}$, we can use relatively tight normal priors because we expect TPR and TNR to be high. Similarly, we can specify a moderately informative beta prior for $\delta_{Q}$ since we do not expect the extra uncertainty in the observed diagnostic viral load below LoQ to be huge. As for the positive-valued parameters $\left\{ \gamma_1, \gamma_2, \kappa_1, \kappa_2 \right \}$, we can use gamma distributions as priors such that the resulting prior distributions for $q_i$ and $w_i$ have reasonable values as suggested by previous literature. 

\subsection{sgRNA viral load imputation} \label{sec_impute}

In a resource-constrained setting, it is common that everyone in a study was tested for diagnostic RNA but only a subset were tested for sgRNA. In this case, our model can be used to impute the sgRNA viral trajectories for people who only had diagnostic RNA viral load and covariates data by treating their sgRNA measurements as missing data. Specifically, we can combine the people without sgRNA samples with the people who had complete data and fit the model to everyone to estimate the posterior distributions of  $\left\{ v'_{p,i}, w'_{a,i}, w'_{b,i} \right \}$ by borrowing information from the rest of the parameters. 

\section{Application to the COVID-19 PEP study} \label{sec_application}

\subsection{Model implementation} \label{sec_fitting}

We applied our methods to analyze data collected from the 80 COVID-19 PEP study participants who had at least 2 positive sgRNA samples during the first 14 days of follow-up (Figures S1-2 of the Supplement). We chose this subset of participants because we are interested in examining the viral dynamics and seroconversion among SARS-CoV-2-infected people who had sustained viral shedding. Among them, 33 participants had at least 1 DBS sample with confirmed results, hence their anti-spike IgG antibody data contributed to the seroconversion part of the model. For our primary model, the following covariates were included as correlates for peak diagnostic viral load ($\bm{X}_{vp}$): age, sex at birth, treatment arm (hydroxychloroquine vs. placebo), and presence of COVID-19 symptoms defined based on the case definition approved by the U.S. Centers for Disease Control and Prevention (CDC) (\citealp{cdc2020}). The peak diagnostic viral load, $v_p$, was included as a correlate for seroconversion probability ($\bm{X}_{C}$). All analysis was implemented in R 4.2.1 (\citealp{r2022}) using JAGS 4.3.0 (\citealp{plummer2003jags}). More details, including prior specification, Markov Chain Monte Carlo (MCMC) setup, and model diagnostics, can be found in Section S1 of the Supplement. 

\subsection{Model results} \label{sec_results}

\begin{figure}[!p]
	\centering
	\includegraphics[width=0.75\textwidth]{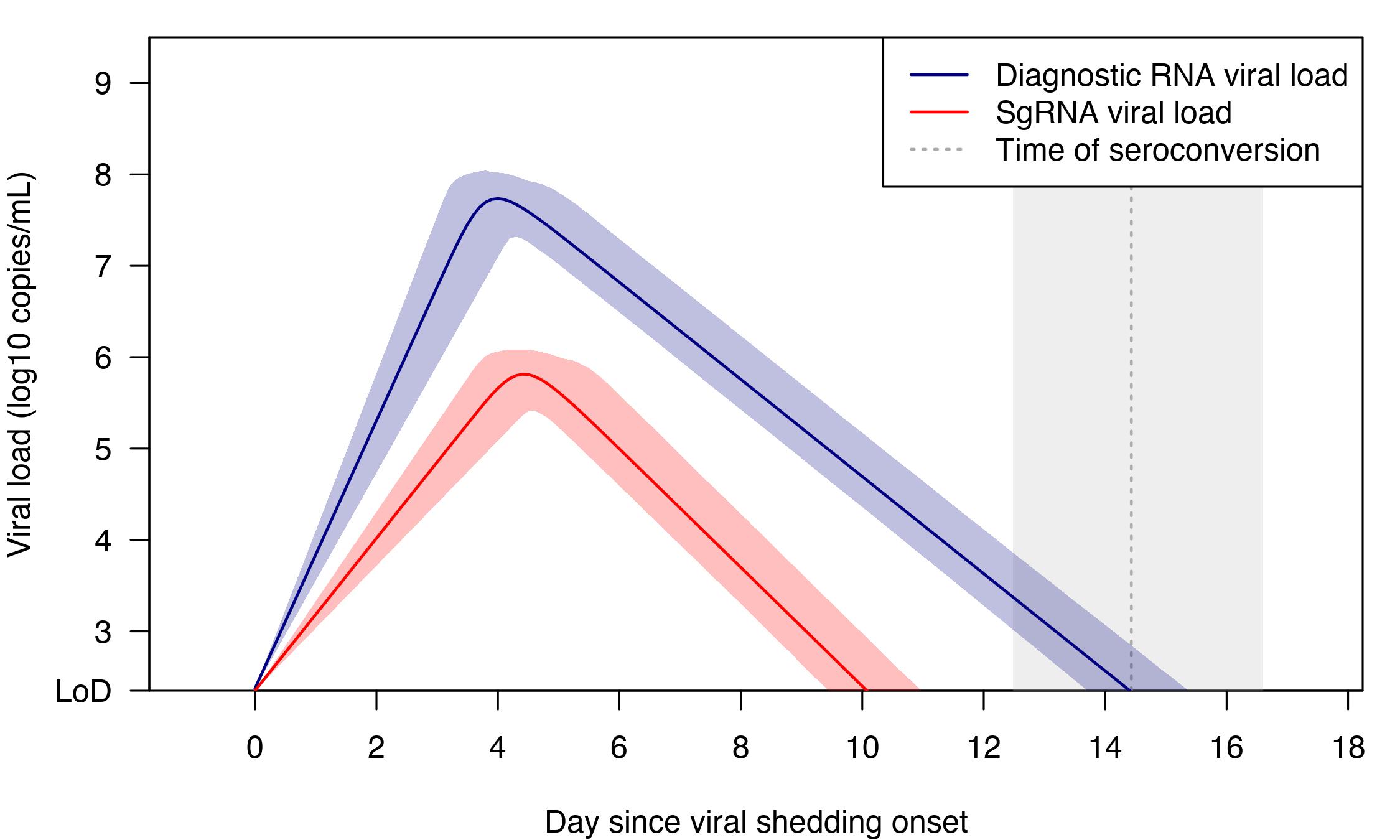}
	\caption{The posterior means and 95\% credible intervals (CI) of the population-level diagnostic and sgRNA viral load trajectories and time of anti-spike IgG seroconversion (among those who seroconverted). }
	\label{fig_vl_traj}
\end{figure}

On average, the diagnostic viral load reaches a peak of $8.0$ (95\% credible interval, CI: $\left[7.8, 8.1\right]$) log10 copies/ml after 3.8 (95\% CI: $\left[3.2, 4.7\right]$) days and then clears after 10.5 (95\% CI: $\left[10.1, 10.9\right]$) days. The sgRNA viral load reaches a peak of $6.0$ (95\% CI: $\left[5.9, 6.2\right]$) log10 copies/ml approximately 0.6 (95\% CI: $\left[0.3, 0.8\right]$) day after the peak diagnostic viral load and then clears after only 5.7 (95\% CI: $\left[5.3, 6.1\right]$) days, approximately 4.3 (95\% CI: $\left[4.0, 4.6\right]$) days before the diagnostic viral clearance (Table \ref{tab_covid_summ} and Figure \ref{fig_vl_traj}). The estimated infectious period, as approximated by the total duration of sgRNA viral shedding, is 10.1 days (95\% CI: $\left[9.4, 11.0\right]$). The posterior means and 95\% CIs of individual viral load trajectories are shown in Figures S1-2 in the Supplement. 

\begin{figure}[!p]
	\centering
	\includegraphics[width=0.8\textwidth]{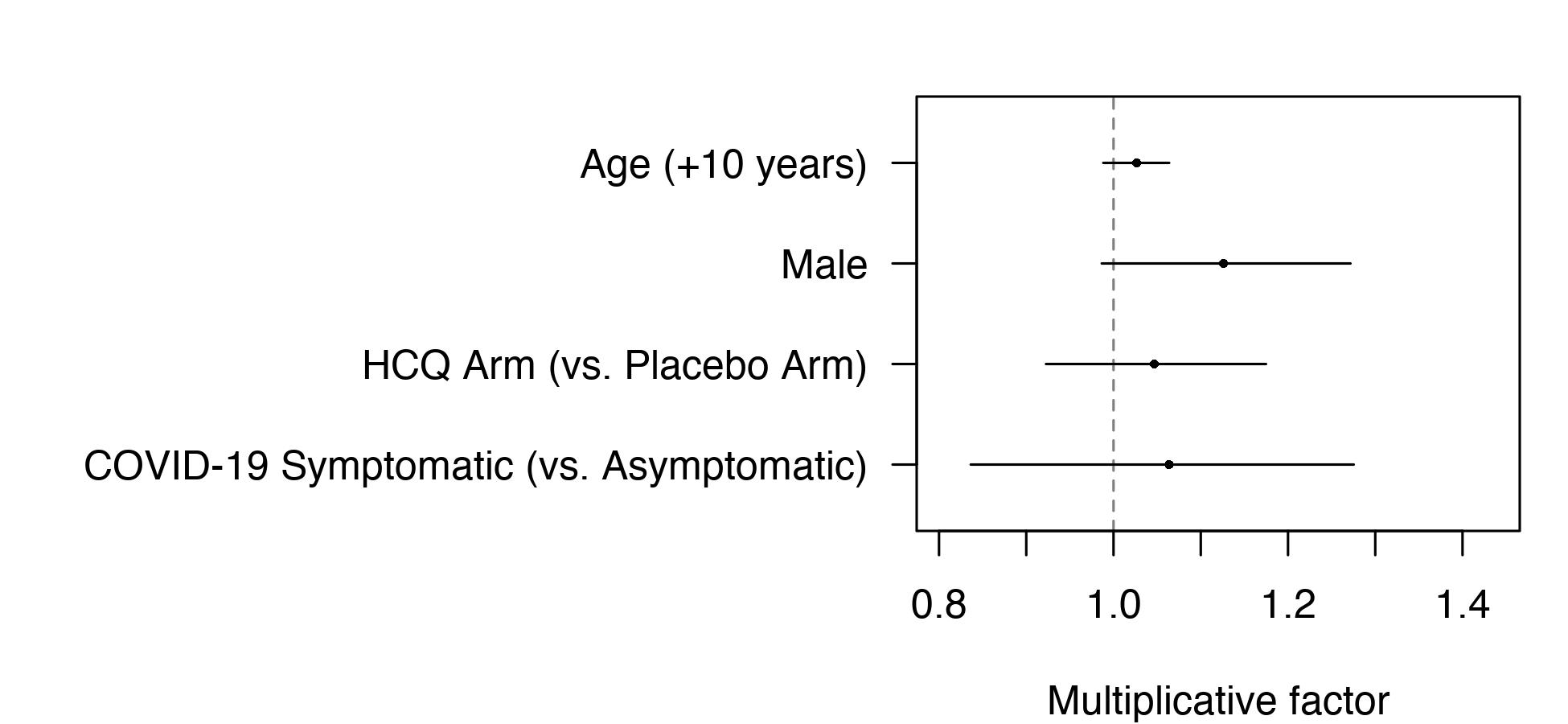}
	\caption{The posterior means and 95\% credible intervals of the multiplicative factors associated with various patient characteristics for an increase in peak diagnostic viral load on the log10 copies/ml scale. }
	\label{fig_covid_summ}
\end{figure}

We estimated the multiplicative factors associated with various patient characteristics for an increase in peak diagnostic viral load on the log10 copies/ml scale (Figure \ref{fig_covid_summ}). The posterior means of the parameters indicated that being older, male, in the hydroxychloroquine arm, and reporting COVID-19 symptoms were positively associated with peak viral load in our data. However, all the posterior 95\% CIs were wide and overlapped with 1, indicating that there was high uncertainty associated with these estimates. 

\begin{figure}[!p]
	\centering
	\includegraphics[width=0.8\textwidth]{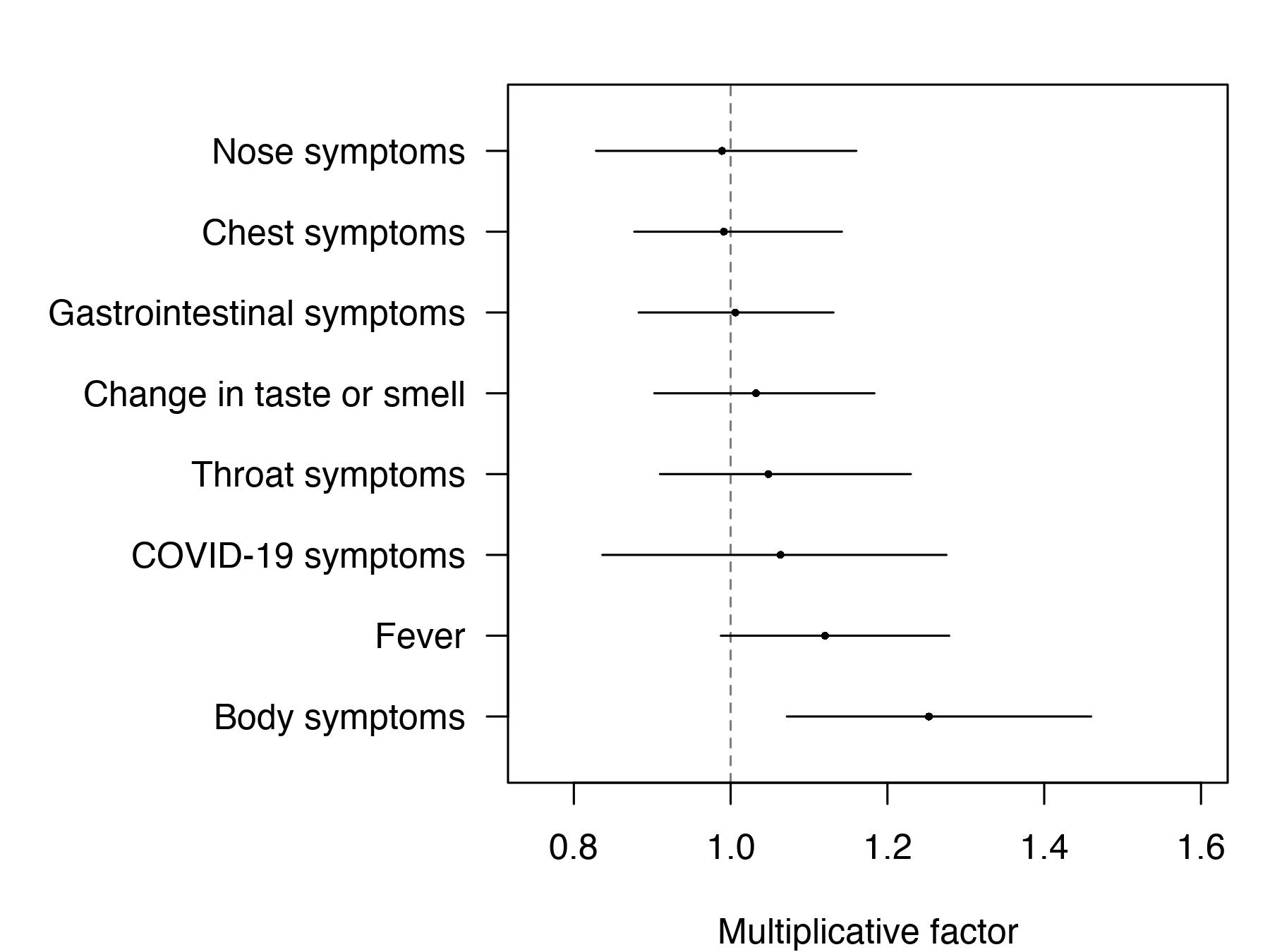}
	\caption{The posterior means and 95\% credible intervals of the multiplicative factors associated with various symptoms for an increase in peak diagnostic viral load on the log10 copies/ml scale. Specifically, ``COVID-19" symptoms were defined based on the case definition approved by the U.S. Centers for Disease Control and Prevention (\citealp{cdc2020}); the rest were defined based on the Flu-PRO patient-reported outcome instrument (\citealp{powers2015development}): ``Chest" symptoms include cough and shortness of breath; ``Taste or Smell" symptoms include olfactory and taste disorders;  ``Nose" symptoms include congestion, runny nose, and olfactory disorder; ``Body" symptoms include chills, myalgia, headache, and fatigue; ``GI" symptoms include diarrhea, nausea, and vomit; ``Throat" symptoms include sore throat and taste disorder. }
	\label{fig_all_summ}
\end{figure}

To identify specific symptoms that might potentially be associated with peak diagnostic viral load, we fitted additional models by replacing the COVID-19 symptoms indicator with other symptom indicators defined based on the Flu-PRO patient-reported outcome instrument (\citealp{powers2015development}), including fever, change in taste or smell, and chest/throat/nose/body/gastrointestinal symptoms. Figure \ref{fig_all_summ} shows the posterior means and 95\% CIs of the multiplicative factors associated with various symptoms for an increase in peak diagnostic viral load on the log10 copies/ml scale. Among all symptoms investigated, having body symptoms (chills, myalgia, headache, or fatigue) was most strongly associated with a higher peak diagnostic viral load (multiplicative factor = 1.25, 95\% CI: $\left[1.07, 1.50 \right]$). That is, on the log10 copies/ml scale, the peak diagnostic RNA viral load was estimated to be 25\% (95\% CI: $\left[7\%, 50\% \right]$) higher among SARS-CoV-2 patients who experienced chills, myalgia, headache, or fatigue than those who did not experience these symptoms. The rest of the symptoms either had close-to-zero or positive estimated associations, but all their 95\% CIs overlapped with 1. 

The overall seroconversion rate was 75.0\% (95\% CI: $\left[62.5\%, 85.0\%\right]$). The peak diagnostic viral load was estimated to be positively associated with the probability of seroconversion (odds ratio = 1.24, 95\% CI: $\left[1.00, 1.54 \right]$, for every 10-fold increase in diagnostic peak viral load). Among those who seroconverted, the mean time from viral shedding onset to the presence of detectable IgG antibodies is 14.4 days (95\% CI: $\left[12.5, 16.6\right]$) (Table \ref{tab_covid_summ} and Figure \ref{fig_vl_traj}). 

\begin{figure}[!p]
	\centering
	\includegraphics[width=0.8\textwidth]{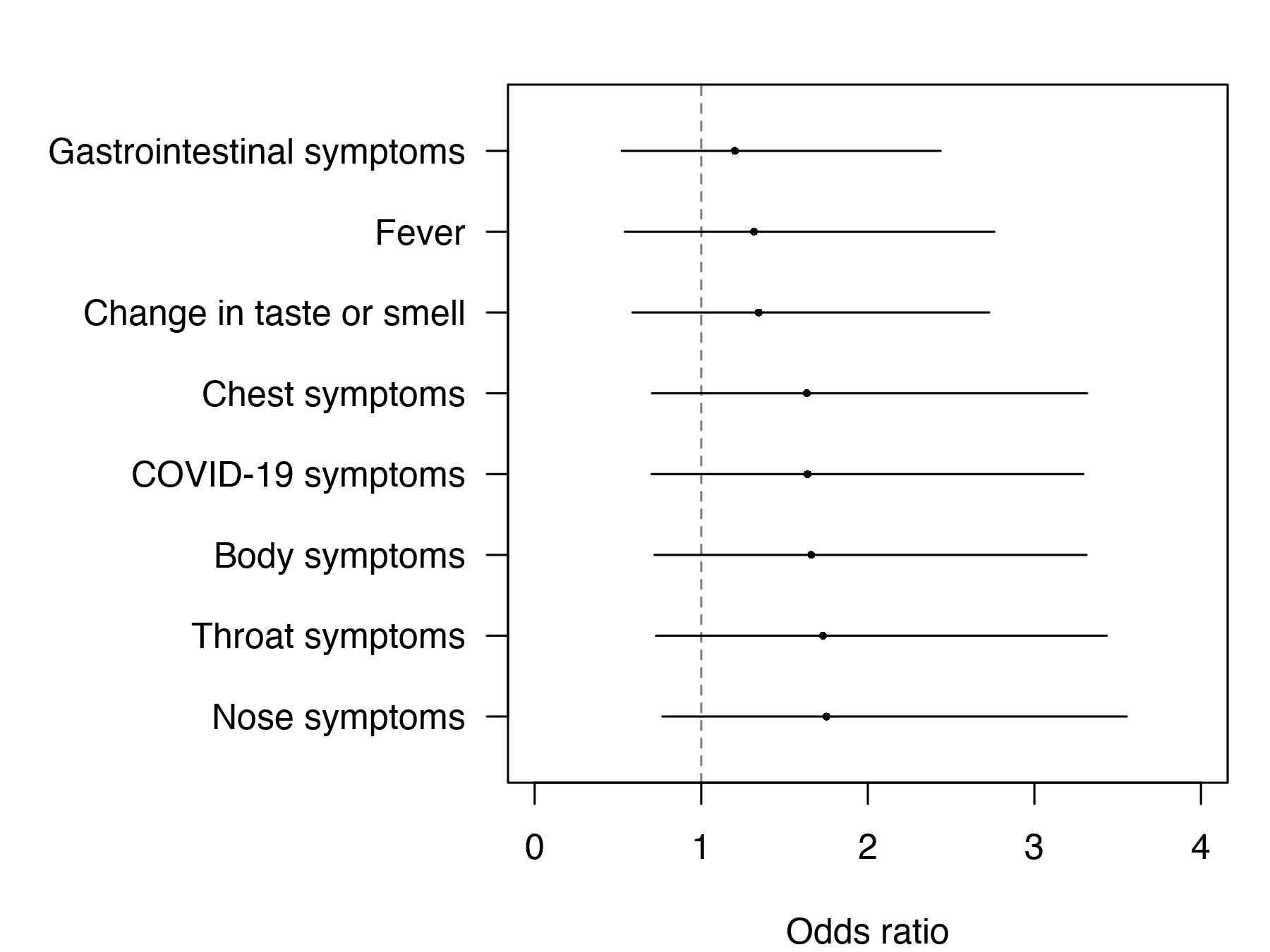}
	\caption{The posterior means and 95\% credible intervals of the odds ratio of seroconversion associated with various symptoms. Specifically, ``COVID-19" symptoms were defined based on the case definition approved by the U.S. Centers for Disease Control and Prevention (\citealp{cdc2020}); ``Chest" symptoms include cough and shortness of breath; ``Taste or Smell" symptoms include olfactory and taste disorders;  ``Nose" symptoms include congestion, runny nose, and olfactory disorder; ``Body" symptoms include chills, myalgia, headache, and fatigue; ``GI" symptoms include diarrhea, nausea, and vomit; ``Throat" symptoms include sore throat and taste disorder. }
	\label{fig_all_summ_sero}
\end{figure}

To identify any symptoms that might potentially be associated with seroconversion rate, we fitted additional models by replacing peak diagnostic viral load with symptom indicators. Figure \ref{fig_all_summ_sero} shows the posterior means and 95\% CIs of the odds ratio of seroconversion associated with various symptoms. All symptoms investigated were estimated to be positively associated with the probability of seroconversion but all posterior 95\% CIs overlapped with 1. 

The probabilities that a sample collected during viral shedding would be tested positive for diagnostic and sgRNA (TPR) are 91.8\% (95\% CI: $\left[90.1\%, 93.3\%\right]$) and 89.6\% (95\% CI: $\left[87.4\%, 91.6\%\right]$) respectively; the probabilities that a sample collected outside the viral shedding periods would be tested negative for diagnostic and sgRNA (TNR) are 99.5\% (95\% CI: $\left[99.4\%, 99.6\%\right]$) and 99.6\% (95\% CI: $\left[99.5\%, 99.7\%\right]$) respectively (Table \ref{tab_covid_summ}). Note that these values account for not only the properties inherent to the RT-PCR tests such as sensitivity and specificity, but also the potential errors that occurred due to misplaced swabs, contaminated samples, or improper swabbing that were common in outpatient household studies. 

Our model results indicated that there were no significant correlations among peak diagnostic viral load, time from shedding onset to peak, and time from peak to diagnostic viral clearance (Table S1 in the Supplement). As for the extra uncertainty in diagnostic viral load measurements below LoQ (104.9 copies/ml, corresponding to a Ct value of 31), our model estimated that the variance in the observed diagnostic viral load ($\sigma^2_y$) was approximately 25\% (95\% CI: $\left[ 13\%, 38\% \right]$) higher if the observed values were below LoQ as compared to those above LoQ. 

\subsection{Cross-validation for sgRNA viral load imputation and seroconversion probability estimation} \label{sec_cv} 

We conducted an 10-fold cross-validation exercise to examine our model's ability to impute sgRNA viral load and seroconversion for people who only have diagnostic viral load and covariate data. Specifically, we randomly divided the 80 participants into 10 groups of 8 people. For each iteration, we removed the sgRNA viral load data (and antibody data if available) from one group of participants and fitted the model using the rest of the observed data. In figures S3-4 in the Supplement, we present the observed and imputed sgRNA viral load trajectories and the estimated probability of seroconversion from all iterations. The posterior means of the sgRNA viral trajectories were very close to the masked sgRNA viral load measurements. As expected, the imputed sgRNA viral load had wider posterior 95\% CIs than those of the diagnostic viral trajectories. The estimated seroconversion probabilities ranged from 64\% to 89\%, a reasonable range given the masked antibody data. 

\section{Discussion}\label{sec_discuss}

In this paper, we presented a joint Bayesian hierarchical model that allows for borrowing of information between viral load and antibody data to make inferences on key characteristics of diagnostic and sgRNA viral dynamics and seroconversion. Our method can be used to identify potential correlates of viral load trajectories and propensity for seroconversion, to estimate the seroconversion time after a SARS-CoV-2 infection, and to impute sgRNA viral load for people who only have diagnostic RNA viral load data. By jointly modeling the diagnostic and sgRNA viral trajectories, our method is able to take into account the dynamical relationship and correlation structure between the two types of viral load. Our model also uses indicator variables to represent latent and observed viral shedding status separately, accounting for false positive and false negative viral detection due to not only the RT-PCR test properties but also the potential sample collection errors in outpatient study settings. As a demonstrating example, we implemented our proposed methods using JAGS in R and analyzed data collected from 80 participants of the COVID-19 PEP study who had at least 2 positive sgRNA samples during the first 14 days of follow-up. A cross-validation exercise was also conducted to examine the model's ability to impute sgRNA viral load trajectories. 

Based on the COVID-19 PEP data, our model-estimated population-level viral load trajectories were characterized by a rapid rise to reach viral peak and subsequent slower decline, similar to what has been demonstrated in previous literature (\citealp{goyal2020potency, neante2020modeling, ke2021vivo, kissler2021viral, kissler2021viral1, SK2021trajectory, jones2021estimating, singanayagam2021community, deming2022Detection}). The model results on diagnostic and sgRNA viral shedding duration were consistent with previous findings that SARS-CoV-2 infectivity usually terminates after 10 days (\citealp{van2021duration, wolfel2020virological, bullard2020predicting, arons2020presymptomatic, covid2020clinical, buder2022contribution}) while SARS-CoV-2 RNA is detectable in respiratory samples for an average of 17 days (\citealp{cevik2021sars}). In addition, our analysis confirms that anti-spike IgG seroconversion is positively associated with viral load (\citealp{liu2021predictors, masia2021sars}). To our knowledge, our model is the first that quantifies the association between \textit{peak} SARS-CoV-2 viral load and IgG seroconversion rate and the time from SARS-CoV-2 \textit{infection} to IgG seroconversion. 

One limitation of our paper is the small sample size of the COVID-19 PEP data especially the limited number of people with seroconversion data. Repeating the analysis with data from more participants and more frequent antibody sampling schedule may increase the precision of the parameter estimates. In addition, we only focused on the people with sustained viral shedding, which we subset based on the criteria of having at least 2 positive sgRNA samples during the first 14 days. Extending our method to include people who only had one positive sample is possible. However, adjustment needs to be made to adequately account for the possible cases of intermittent viral detection at the very end of viral shedding, clearance of the virus by innate immunity due to low viral inoculum, or/and limited infections attributed to immune priming by prior seasonal coronaviruses (\citealp{SK2021trajectory}). 

In addition to identifying correlates of peak diagnostic viral load, as demonstrated in the analysis of PEP data, our model can be used to investigate factors associated with other viral dynamics characteristics as well. For example, we can estimate the potential association between vaccination and viral shedding duration by using vaccination status as a covariate for time from viral shedding onset to peak ($\bm{X}_{wa,i}$) and/or time from peak to viral clearance ($\bm{X}_{wb,i}$). Although we only focused on anti-spike IgG seroconversion in this paper, our model can be generalized to study other immune responses, such as the development of anti-nucleocapsid IgG and Immunoglobulin M (IgM) antibodies, if relevant data is available. Given the flexible structure of the joint hierarchical model, it is straight forward to extend our methods to model multiple immune markers simultaneously. 

Our model will be a valuable tool for future SARS-CoV-2 vaccine studies that focus on assessing the vaccine effect on transmission (VET)(\citealp{kennedy2021estimating, follmann2022vaccine}). Since detection of sgRNA indicates active viral replication and tracks infectious virus, our methods can be used to impute sgRNA viral load and better estimate proxies of VET. Our model can also be used to better understand the potential impact of interventions that reduce viral load on seroconversion rate. For example, previous vaccine studies have shown that vaccinated people tended to have lower viral load at diagnosis and have lower rate of seroconversion as compared to the control group (\citealp{pajon2022initial, follmann2022antinucleocapsid}). Models like ours could help explain how much of the decrease in seroconversion rates is due to vaccination and how much is due to the vaccination effect on viral load. Finally, our proposed model can facilitate analysis of household transmission studies by providing estimated viral load trajectories for each individual in a household to inform the direction of transmission. 

\section{Software} \label{sec_software}

Software in the form of R code and JAGS code is available online at \url{https://github.com/dq0708/joint_vl_sero}.

\section{Supplementary Material} \label{sec_supp}

Supplementary material is available online at \url{http://biostatistics.oxfordjournals.org}.

\section*{Acknowledgments}

Funding for the project was provided by Bill \& Melinda Gates Foundation. The authors thank the Hydroxychloroquine COVID-19 PEP Study Team and participants for providing scientific insights and data. Special thanks to Dr. Ruanne Barnabas and Dr. Anna Bershteyn for reviewing the manuscript drafts and providing valuable suggestions. We also thank Dr. Leigh Fisher for additional advice regarding RT-PCR viral load quantification. 

{\it Conflict of Interest}: None declared.

\bibliographystyle{biorefs}
\bibliography{References}

\clearpage


\begin{table}[!p]
\caption{The posterior means and 95\% credible intervals (CI) for selected quantities of interest. In particular, $n = 80$ is the total number of people with diagnostic RNA viral load data. }
\label{tab_covid_summ}
\begin{adjustwidth}{-2.2cm}{0cm}
\renewcommand{\arraystretch}{1.5}
\begin{tabular}{llll}
\hline
Type & Quantity of interest & Estimand & Posterior mean [95\% CI] \\ \hline
Diagnostic RNA & Mean time from shedding onset to peak viral load (days) & $\sum_{i = 1}^n \hat{w}_{a,i}/n$ & $3.8 \left[3.2 - 4.7 \right]$ \\
viral load & Mean time from peak viral load to viral clearance (days) & $\sum_{i = 1}^n \hat{w}_{b,i}/n$ & $10.5 \left[10.1 - 10.9 \right]$ \\
& Mean peak viral load (log10 copies/ml)& $\sum_{i = 1}^n \hat{v}_{p,i}/n + \text{LoD}$ & $8.0 \left[7.8 - 8.1 \right]$ \\ 
& True Positive Rate & $\expit(\hat{\alpha}_0 + \hat{\alpha}_1)$ & $91.8\% \left[ 90.1\% - 93.3\% \right]$ \\
& True Negative Rate & $1 - \expit(\hat{\alpha}_0)$ & $99.5\% \left[99.4\% - 99.6\% \right]$ \\ \hline
sgRNA & Mean time from shedding onset to peak viral load (days) & $\sum_{i = 1}^n \hat{w}'_{a,i}/n$ & $4.4 \left[3.8 - 5.3 \right]$ \\
viral load & Mean time from peak viral load to viral clearance (days) & $\sum_{i = 1}^n \hat{w}'_{b,i}/n$ & $5.7 \left[5.3 - 6.1 \right]$ \\
& Mean peak viral load (log10 copies/ml) & $\sum_{i = 1}^n \hat{v}'_{p,i}/n + \text{LoD}'$ & $6.0 \left[5.9 - 6.2 \right]$ \\ 
& True Positive Rate &  $\expit(\hat{\alpha}'_0 + \hat{\alpha}'_1)$ & $89.6\% \left[ 87.4\% - 91.6\% \right]$ \\
& True Negative Rate & $1 - \expit(\hat{\alpha}'_0)$ & $99.6\% \left[99.5\% - 99.7\% \right]$ \\ \hline
Anti-spike IgG & The overall seroconversion rate & $\sum_{i = 1}^n \hat{C}_{i}/n$ & $75.0\% \left[62.5\% - 85.0\% \right]$  \\
Seroconversion & The odds ratio of seroconversion associated with & $\exp \left(\hat{\bm{\beta}}_{C}[2]\right)$ & $1.24 \left[1.00 - 1.54 \right]$  \\
& \quad every 10-fold increase in peak diagnostic viral copies & &  \\
& Mean time from infection to seroconversion (days) & $\hat{\kappa_1}/\hat{\kappa_2} $ & $14.4 \left[12.5 - 16.6 \right]$ \\
\hline 
\end{tabular}
\end{adjustwidth}
\end{table}

\clearpage

\end{document}